**Magnetic and magnetodielectric coupling anomalies in the Haldane spin-chain system $Nd_2BaNiO_5$**

Tathamay Basu[1], Niharika Mohapatra[2], Kiran Singh[3] and E. V. Sampathkumaran[1]

[1]*Tata Institute of Fundamental Research, Homi Bhabha Road, Colaba, Mumbai-400005, India*

[2]*Indian Institute of Technology Bhubaneswar, Bhubaneswar, Odisha- 171013, India*

[3]*UGC-DAE Consortium for Scientific Research, Indore Centre, Khandwa Road, Indore - 452001, India*

We report the magnetic, heat-capacity, dielectric and magnetodielectric (MDE) behaviour of a Haldane spin-chain compound containing light rare-earth ion, $Nd_2BaNiO_5$, in detail, as a function of temperature ($T$) and magnetic field ($H$) down to 2 K. In addition to the well-known long range antiferromagnetic order setting in at ($T_N$=) 48 K as indicated in dc magnetization ($M$), we have observed another magnetic transition near 10 K; this transition appears to be of a glassy-type which vanishes with a marginal application of external magnetic field (even $H$= 100 Oe). There are corresponding anomalies in dielectric constant ($\varepsilon'$) as well with variation of $T$. The isothermal $M(H)$ curves at 2 and 5 K reveal the existence of a magnetic-field induced transition around 90 kOe; the isothermal $\varepsilon'(H)$ also tracks such a metamagnetic transition. These results illustrate the MDE coupling in this compound. Additionally, we observe a strong frequency dependence of a step in $\varepsilon'(T)$ with this feature appearing around 25-30 K for the lowest frequency of 1 kHz, far below $T_N$. This is attributed to interplay between crystal-field effect and exchange interaction between Nd and Ni, which establishes the sensitivity of dielectric measurements to detect such effects. Interestingly enough, the observed dispersions of the $\varepsilon'(T)$ curves is essentially $H$-independent in the entire $T$-range of measurement, despite the existence of MDE coupling, which is in sharp contrast with other heavy rare-earth members in this series.

The nickelates of the type $R_2BaNiO_5$ (R=Y, rare-earths) have long been thought of as model systems for studying one-dimensional magnetic behavior of integer-spins as well as to understand dimensional crossover with varying temperature (*T*). This family of oxides is particularly interesting [1] due to the existence of Haldane spin-gap even in the magnetically ordered state in the case of magnetic-moment containing rare-earths [see, for instance, Refs 2 and 3]. Recently, we carried out exhaustive investigations [4-8] on many heavy rare-earth members of this family (*R*= Gd, Dy, and Er), bringing out new interesting properties, for example, (i) an additional magnetic transition at temperatures far below respective Néel temperature ($T_N$), (ii) magnetoelectric coupling, and (iii) not so-commonly known 're-entrant multiglass-like' behavior, that too of unusual types [5,7]. We also brought out magnetic-field-induced metamagnetic transitions and associated magnetodielectric (MDE) coupling effects. We therefore considered it worthwhile to subject one of the light rare-earth members, e.g., $Nd_2BaNiO_5$, for similar investigations for exploration of these anomalies in light rare-earth members of this family.

The compound $Nd_2BaNiO_5$ forms in an orthorhombic structure (space group *Immm*) consisting of chains of flattened $NiO_6$ octahedra along 'a' axis, separated by Ba and Nd along 'c'. At 1.5 K, the magnetic moment of Nd is nearly aligned along c-axis, while that of $Ni^{2+}$ forms a small angle with this axis [3]. The $Ni^{+2}$ and $Nd^{+3}$ moments are coupled antiferromagnetically along the chain ('a' axis) but coupled ferromagnetically along 'b' axis. The Ni moments start to order antiferromagnetically below 48 K [9], and reaches its saturation value around 35 K with decreasing *T*. The Nd moments also orders antiferromagnetically below 48 K, but the magnitude of its moment increases relatively slowly with decreasing *T* [3]. The magnetic susceptibility ($\chi$) shows a broad maximum around 26 K which was attributed to the depletion of the higher-lying level of the Kramers doublet, split by Nd-Ni magnetic interactions [10]. Inelastic neutron scattering studies revealed clear evidence for one-dimensional gap excitations [11].

Polycrystalline compound $Nd_2BaNiO_5$ was prepared by solid state preparation as described in Ref. 1 starting with stoichiometric amounts of respective high-purity (>99.95%) oxides, $Nd_2O_3$, NiO and $BaCO_3$. The formation of the sample with reported structure was confirmed by x-ray diffraction. The *dc* and *ac* magnetization (*M)* were measured with the help of a Superconducting Quantum Interference Device (Quantum Design, QD) and isothermal *M* was carried out using Vibrating Sample Magnetometer (QD). Heat-capacity (C) measurements were performed with a commercial (QD) Physical Property Measurements System (PPMS). The complex dielectric measurements were performed using a LCR meter (Agilent E4980A) with a homemade sample holder coupled to the PPMS.

Temperature dependence of *dc M* obtained for zero-field-cooled (*ZFC,* from 150 K) and field-cooled (*FC*) conditions under the application of 100Oe as well as in 5 kOe is shown in figure 1. As well-known in the literature [3], the feature at $T_N$ is very weak. The broad maximum around 27 K has been attributed to the crystal field effect of $Nd^{+3}$ [3, 10]. As the temperature is lowered, the ZFC-FC curves for *H*= 100 Oe bifurcate near 10 K, with the ZFC-curve showing an upturn and FC-curve showing a downturn. Similar bifurcation of ZFC-FC curves (attributed to spin-glass freezing) has been reported by Popova et al [12]. The corresponding curves for *H*= 5 kOe exhibit an upturn near 10 K, however without any bifurcation. It is therefore clear that there is another magnetic feature at a temperature (<10 K) lower than $T_N$. The fact that the bifurcation of the ZFC-FC curves is suppressed by a small application of external magnetic field is in favour of spin-glass freezing. If one sees the nature of ZFC-FC bifurcation, FC curve does not remain nearly constant with lowering *T* as expected for spin-glass freezing, that is, there is a monotonic increase. Therefore there is a possibility that a para/ferromagnetic component also coexists at such low temperatures.

We have performed ac $\chi$ measurements (with $H_{ac}$= 1 Oe) to confirm this scenario, the results of which are shown in figure 2. The nature of the curve measured in the absence of an external dc *H* for the real part ($\chi'$) above 10 K is similar to that observed in dc $\chi$ data; however, a peak appears at

~10K for a frequency (ν) of 1.3 Hz, and the curve around this *T*-region seems to shift to a marginally higher temperature as ν is increased, say, towards 1339 Hz. There is a corresponding anomaly in the imaginary part ($\chi''$) as well. An application of a small dc magnetic field (e.g., 100 Oe) completely suppresses these features (the peak around 10 K and ν-dependence). These are characteristic features of spin-glass freezing, thereby confirming 're-entrant' magnetism in this compound. We have also measured isothermal remnant magnetization at various temperatures. It was found that remnant magnetization decays gradually over a long period of time below 10 K only, supporting the conclusion on spin-glass freezing (see, Supplementary Material [Ref. 13]).

The results of heat-capacity studies measured in the presence of a few externally applied magnetic field are shown in figure 3, in the form of a plot of *C(T)/T versus T*. While the inset of figure 3b shows the overall profile of the behavior in zero-field in the entire *T*-range of investigation, figure 3a and figure 3b show the behavior of the curves in an expanded form in narrow temperature ranges to highlight the features. While we see a weak peak at $T_N$ (see figure 3b) in zero field, there is no clear-cut peak around 10 K (see figure 3a), which is consistent with glassy nature of the magnetism around 10 K. A careful look at the curves taken in the presence of external magnetic fields, measured with *H* up to 140 kOe, yields further insight. The weak peak at $T_N$ undergoes a marginal suppression towards lower temperature with increasing *H* (figure 3b), which is consistent with antiferromagnetism of this compound. There is also a gradual reduction in absolute values of *C/T* with *H* above 12 K*;* interestingly enough, this tendency is reversed below 12 K, and the values tend to increase with increasing *H* (figure 3a). This sign crossover of a net change in *C* indicating an enhancement of entropy with applied field emphasizes a switchover to a more disordered magnetic structure with decreasing temperature, thus supporting re-entrant magnetism.

The compound is highly insulating as measured by electrical resistivity; room temperature value of resistivity is several MΩ-cm, increasing with decreasing temperature. The value is several GΩ-cm around 100 K and could not be measured below 80 K because of the limitation of the instrument (electrometer). The complex dielectric property was measured as a function of $T$ during heating (warming rate 0.5 K/min) with different ν values (1-100 kHz). A noteworthy point is that the dielectric constant (figure 4a) exhibits a sudden upturn in the range 25-35 K with the loss factor (tanδ) showing a corresponding feature (that is, a peak at ~ 30 K for ν= 1 kHz, see figure 4b). It is to be noted that these features appear at a temperature range far below $T_N$. We attribute this to an interplay between crystal-field-effect and magnetic interaction as mentioned at the introduction. This result reveals the sensitivity of complex dielectric measurements of such effects. Additionally, this feature shows a huge ν-dependence; for instance, the upturn in ε' shifts to 35-50 K and the peak in tanδ appearing at ~50 K for ν= 100 kHz. An additional upturn occurs around 50 K both in ε' and tanδ which could be related to the onset of magnetic ordering. With further increase of temperature, another frequency dependent feature (around 80 K, clear in tanδ plots) appears which could be due to the decrease of insulating behaviour.

An observation that is being stressed is that there is a weak feature (a change in slope or drop) in ε′ at much lower temperatures, if one expands the curves below 15 K, as shown in the inset of figure 4a. This exhibits a small ν-dependence. Considering that the χ data indicate a magnetic anomaly around the same temperature region, this may be magnetic in its origin; which illustrates the coupling between magnetic and electric state. The low values of tanδ confirm the highly insulating nature of this sample in the $T$-region of interest which rules out any extrinsic contributions.

In order to prove further MDE coupling, we have performed isothermal $M(H)$ and ε'($H$) experiments with the rate of variation of $H$ of 70 Oe/s at different $T$. It is known from single crystal studies [14] that there is a field-induced metamagnetic transition around 90 kOe for $T$<<10K. We are able to observe a sudden increase of the slope of the $M(H)$ curve below 10 K even in polycrystalline sample (see figure 5a and b for the data at 2 and 5 K), as a signature of this metamagnetic transition. This $M(H)$ curve also shows a small hysteresis around 70 kOe at 2 K (not shown), which suggests disorder-broadened first-order nature of this transition. However, the $M(H)$ plot at 20 K does not exhibit such a feature (not shown here), varying essentially linearly up to 140 kOe. This trend in $M(H)$ curve is consistent with a change in magnetic state across 10 K, proposed above. The $H$-dependent dielectric constant at 2 K (see figure 5c for ν= 50 kHz) also exhibits an anomaly in the form of a dip followed by an upturn around 90 kOe. There is another feature (dip) around 25 kOe, which is not seen clearly in $M(H)$. The similar results are also observed in $Dy_2BaNiO_5$ where field induced transitions are seen more clearly in ε'(H) [see Ref. 4 and discussion there in]. This implies that dielectric studies could be more revealing to subtle changes in magnetoelectric properties, though it is difficult to pinpoint its exact origin at present. The $H$-dependent dielectric constant at 5 K (figure 5d) also shows a clear change around 90 kOe due to metamagnetic transition. This result establishes MDE coupling in this compound. We have confirmed this by measuring ε'($H$) at some other selective frequencies (not shown here).

Finally, we would like stress that an application of a magnetic field (measured with 10, 30, 50, 80 and 140 kOe) does not have any influence of dielectric behavior observed in zero field. To demonstrate this, we show the data for 140 kOe in figures 5c and 5d. This robustness of electric dipoles to external applications of $H$ is in sharp contrast to that seen in ac χ data, presented above, despite of MDE Coupling. In fact, the curves obtained at several magnetic fields for a given frequency (not shown here) are found to overlap in a broad sense, despite a very small change in the value. This situation is completely different from that observed for heavy rare-earth members [5, 7, 8]. In case of heavy rare-earth members in this series [see Fig.3 in Ref. 5, 7 and 8 for Dy, Er and Gd respectively], the dielectric feature at low $T$ suppresses under application of high magnetic field, whereas, the features remains same up to an application of as high as 140 kOe for the light rare-earth member $Nd_2BaNiO_5$ (see inset of figure 5a and 5c for comparison). This result reveals that there is a significant role of the rare-earth member on the feature of MDE coupling. It is not clear at this moment whether it is due to weaker nature of the MDE coupling with respect to that in heavy rare-earth members of this series. It is also possible that magnetic and electric dipole dynamics are different, despite exhibiting MDE coupling. The reason is that similar inferences have made by us in some other families as well in recent years [15-17], where the magnetoelectric coupling is strong.

We have tried to look for ferroelectricity observed for heavy rare-earth members [4,6,8]; however, we could not detect the same (that is, remnant polarization and its reversibility of sign with opposite poling by electric field by measuring pyroelectric current) in this compound. We wonder whether there is any additional factor playing a role, that is, the radial extension/contraction of 4f orbital, to create the local distortion of Ni-O responsible for the ferroelectricity in this series [6]. The different amount of distortion of $NiO_6$ octahedra in different rare-earth members [18] supports our conclusion on this aspect.

We have reported a magnetic anomaly at low T (<10 K) of a glassy type, in addition to the one at $T_N$ (48 K), in $Nd_2BaNiO_5$, thereby suggesting 're-entrant' magnetic behavior of this compound. The presence of magnetodielectric coupling in this compound is also established. This study thus establishes that these two observations are not restricted to heavy rare-earth members, but appear to a characteristic property of this family of oxides. The dielectric constant interestingly tracks the magnetic feature due to the depopulation of exchange-interaction-split Kramers doublet and its frequency-dependence suggests an interesting slow dynamics of this phenomenon. The dielectric features under an initial application of magnetic field are robust in $Nd_2BaNiO_5$, in sharp contrast with heavy rare-earth members (Dy, Er, Gd) in this series; this indicates an important role of rare-earth ion on the nature of cross-coupling between spin and dipole. A comparison of frequency dependence of ac susceptibility and dielectric constants in the presence of external magnetic fields supports our earlier proposals [15, 16] that the dynamics of electric and magnetic dipoles appear different, despite existence of magnetodielectric coupling, warranting microscopic and theoretical studies on this aspect in this area of great current interest.

**Acknowledgement:**

The authors would like to thank Kartik K. Iyer for his help during measurements.

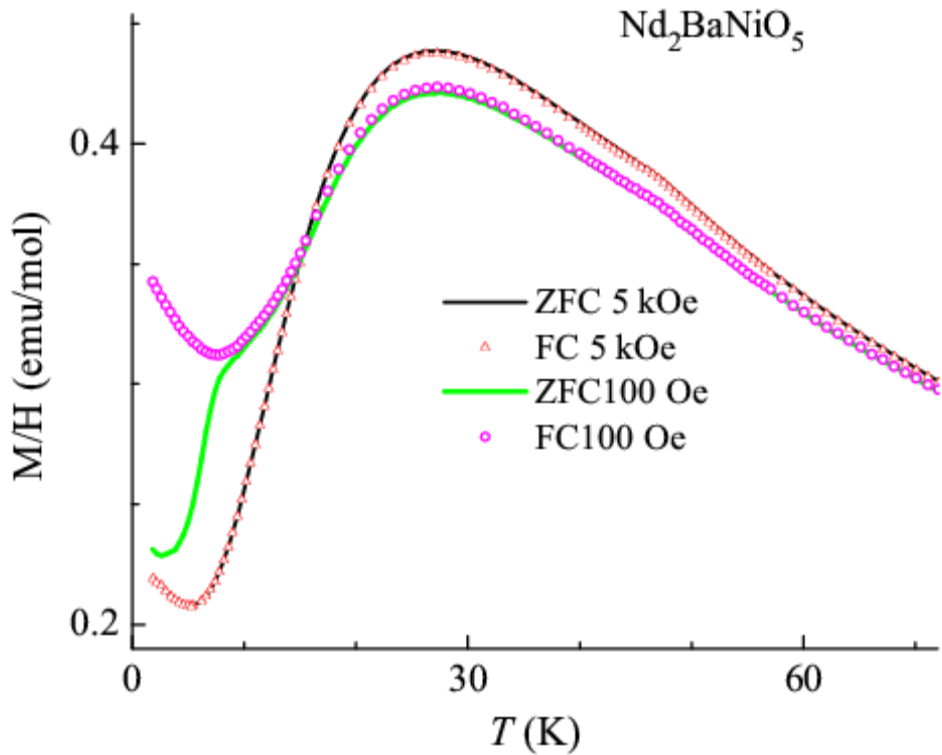

FIG 1. Dc magnetic susceptibility (*M/H*) measured in the presence of 100 Oe and 5 kOe as a function of temperature for *ZFC* and *FC* conditions for $Nd_2BaNiO_5$.

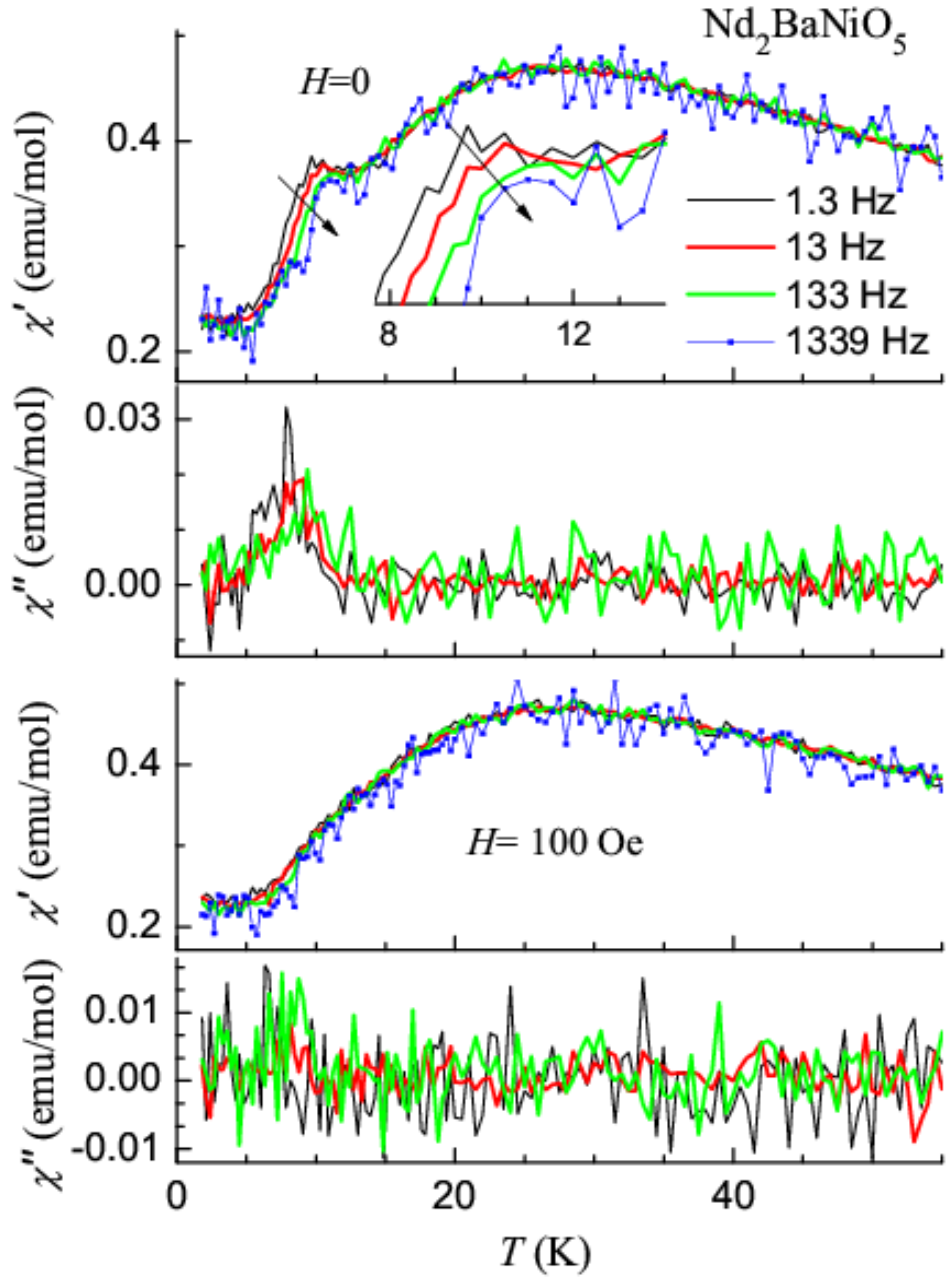

FIG. 2. Real part (χ') and imaginary part (χ") of ac susceptibility as a function of temperature (2 - 55 K) measured with various frequencies (ν = 1.3, 13, 133, 1339 Hz) in the absenace of an external magnetic field as well as in 100 Oe, for $Nd_2BaNiO_5$. The data is noisy for the highest frequency used here and hence deleted for the clarity in χ" plot.

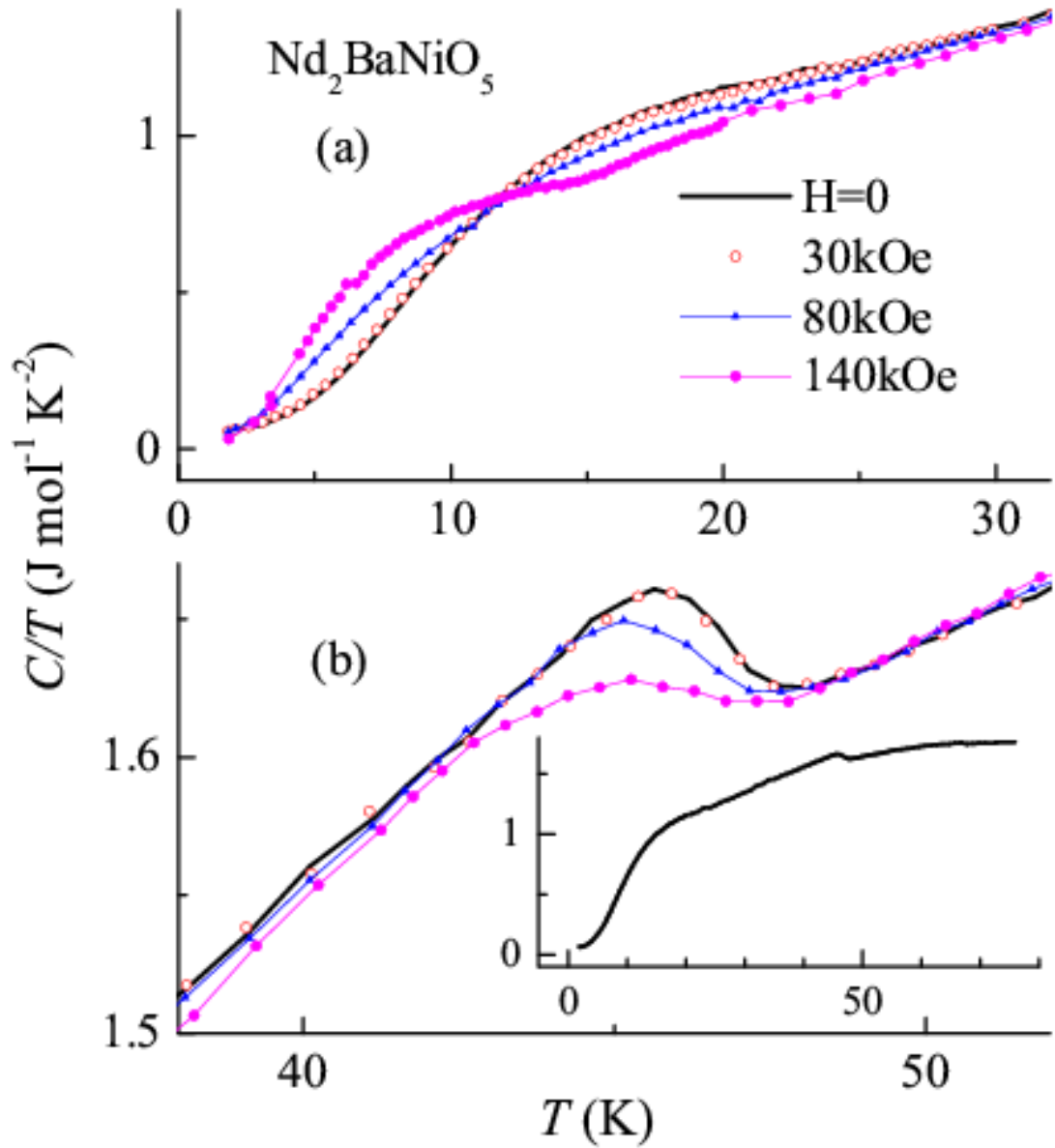

FIG. 3. Heat-capacity divided by Temperature as a function of temperature for $Nd_2BaNiO_5$ in the range **(a)** 2-30 K and **(b)** 38 - 52 K, in zero field as well as in 30, 80 and 140 kOe. In the inset, the curve in the entire *T*-range of measurement in the absence of an external field is shown.

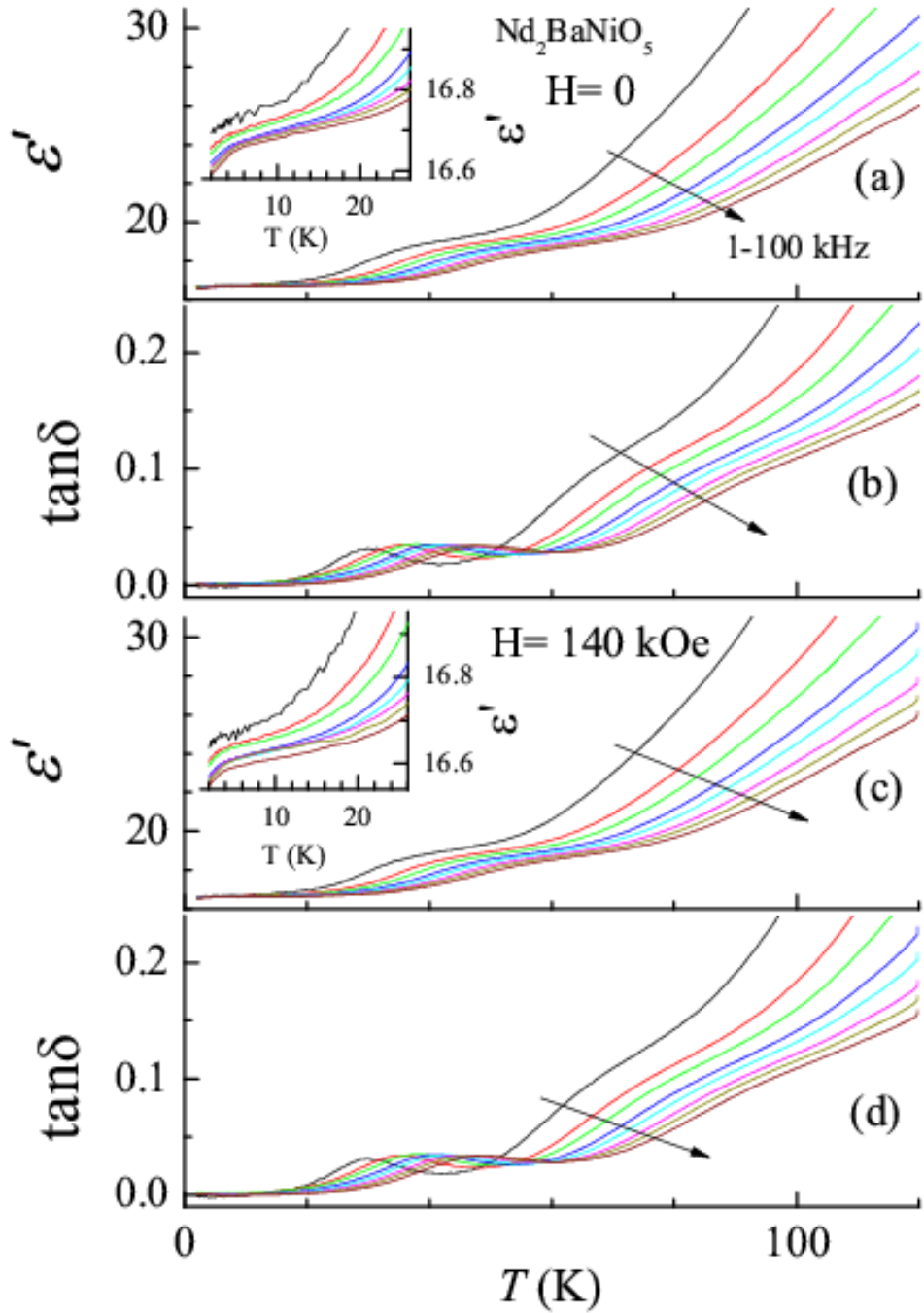

FIG. 4. Frequency dependence of dielectric constant (ε') and loss part (tanδ) as a function of temperature for $Nd_2BaNiO_5$ **(a,b)** in the absence of an external magnetic field and **(c,d)** in 140 kOe. The arrows are drawn to show how the curves move with increasing frequency (1, 5, 10, 20, 30, 50, 70, 100 kHz ). Inset in (a) and (c) is the zoom out plot of ε' vs T in low T-region for H=0 and 140 kOe respectively.

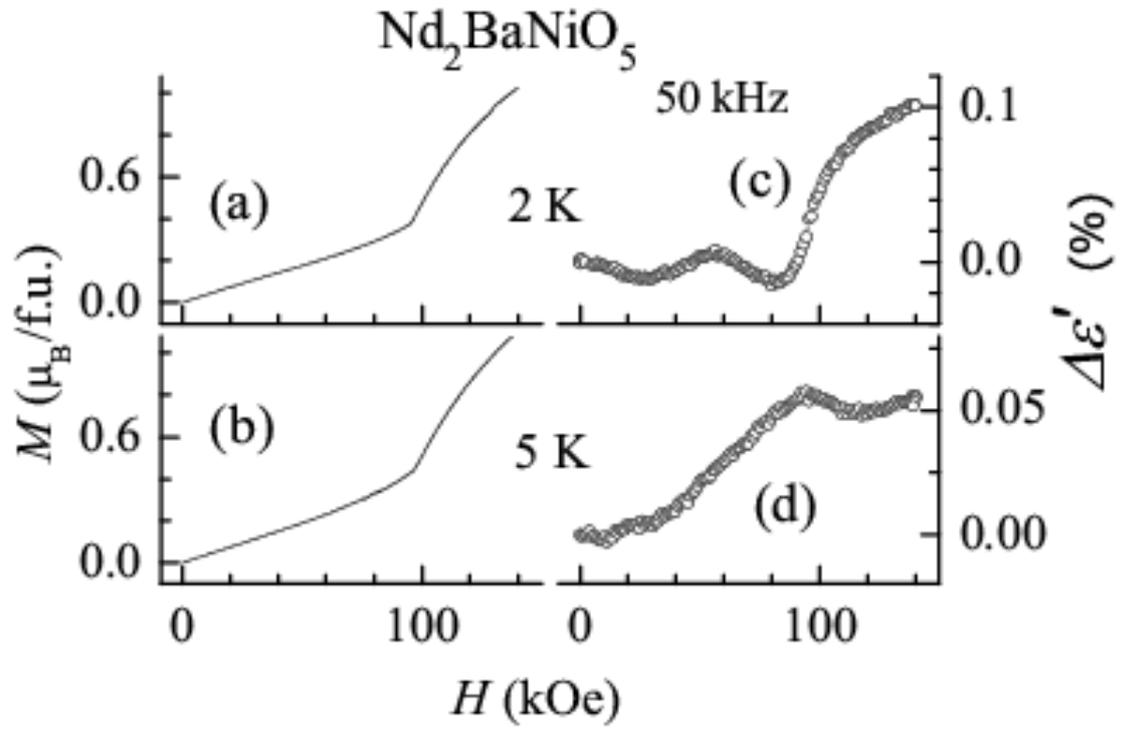

FIG. 5. **(a,b)** Isothermal magnetization and **(c,d)** the fractional change in dielectric constant as a function of magnetic field at 2 and 5 K respectively, for $Nd_2BaNiO_5$. Here, $\Delta\varepsilon' = (\varepsilon'_H - \varepsilon'_{H=0})/\varepsilon'_{H=0}$.

**Magnetic and magnetodielectric coupling anomalies in the Haldane Spin-chain System Nd$_2$BaNiO$_5$**

Tathamay Basu[1], Niharika Mohapatra[2], Kiran Singh[3] and E. V. Sampathkumaran[1]

[1]*Tata Institute of Fundamental Research, Homi Bhabha Road, Colaba, Mumbai-400005, India*

[2]*Indian Institute of Technology Bhubaneswar, Bhubaneswar, Odisha- 171013, India*

[3]*UGC-DAE Consortium for Scientific Research, Indore Centre, Khandwa Road, Indore - 452001, India*

In this supplemental, we have discussed the result of isothermal remnant magnetization (see figure S1) in order to confirm a distinct change in magnetism across 10 K.

We have performed isothermal remnant magnetization ($M_{IRM}$) measurements as a function of time at various temperatures (2, 4, 6, 8 and 10 and 15 K). For this purpose, we cooled the sample from 150 K to the desired temperature, and switched on a field of 5 kOe, which was then switched off after 5 mins. $M_{IRM}$ was then measured as a function of time (*t*). It is distinctly clear from figure S1 that there is a qualitative change in the nature of the curve at 10 K, with respect to the ones at lower temperatures. That is, the $M_{IRM}$, decreases gradually with increasing *t* for *T*<10K, typical of glassy behavior, while at 10 K, the value appears to undergo an increase, the origin of which is not clear. We have also fitted the data below 10 K to a stretched exponential form ($M_{IRM}(t)$= A + B exp[-t/τ]$^{0.5}$, where A and B constants and τ is the relaxation time). The relaxation time τ increases (i.e. the slope become steeper) with initial lowering of *T* below 10 K, consistent with spin-glass behavior; however it again decreases with further lowering of *T* below 5 K which is unusual. This could be due to the coexistence of another magnetic component at such low $T$ as speculated from ZFC-FC dc magnetization data in the manuscript.

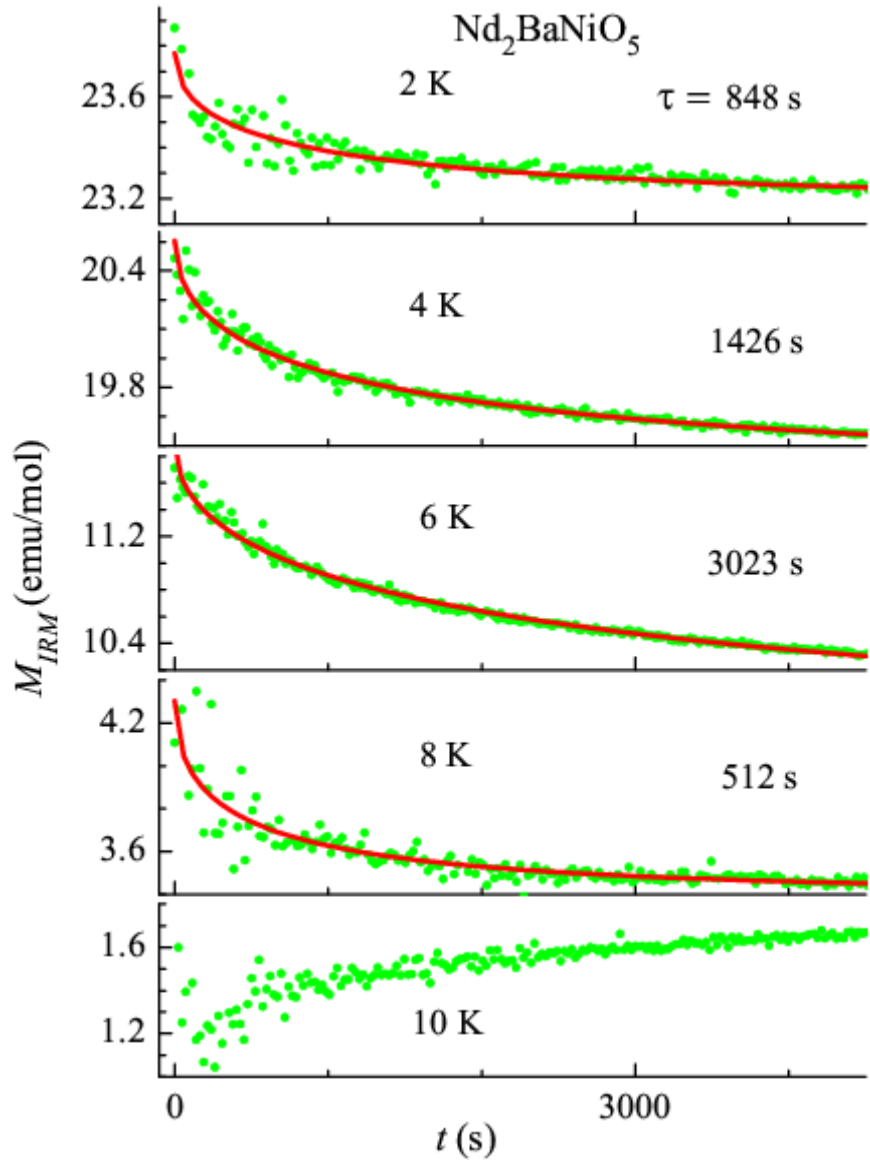

FIG. S1. Isothermal remnant magnetization as a function of time, measured as described in the text, for several temperatures for $Nd_2BaNiO_5$.